\documentclass[twocolumn]{aastex62}

\usepackage{booktabs}

\received{2018}
\revised{2018}
\accepted{2018}
\submitjournal{ApJ}
\shorttitle{KIC\,2568888}
\shortauthors{Theme\ss l et al.}

\begin{document}

\title{KIC\,2568888: To be or not to be a binary}

\correspondingauthor{N. Theme\ss l}
\email{themessl@mps.mpg.de}

\author{N. Theme\ss l}
\affil{Max Planck Institute for Solar System Research, Justus-von-Liebig-Weg 3, D-37077 G{\"o}ttingen, Germany}
\affiliation{Stellar Astrophysics Centre, Department of Physics and Astronomy, Aarhus University, Ny Munkegade 120, DK-8000 Aarhus C, Denmark}
\affiliation{Institute for Astrophysics, Georg-August University G{\"o}ttingen, Friedrich-Hund-Platz 1, D-37077 G{\"o}ttingen, Germany}

\author{S. Hekker}
\affiliation{Max Planck Institute for Solar System Research, Justus-von-Liebig-Weg 3, D-37077 G{\"o}ttingen, Germany}
\affiliation{Stellar Astrophysics Centre, Department of Physics and Astronomy, Aarhus University, Ny Munkegade 120, DK-8000 Aarhus C, Denmark}

\author{A. Mints}
\affiliation{Max Planck Institute for Solar System Research, Justus-von-Liebig-Weg 3, D-37077 G{\"o}ttingen, Germany}
\affiliation{Stellar Astrophysics Centre, Department of Physics and Astronomy, Aarhus University, Ny Munkegade 120, DK-8000 Aarhus C, Denmark}

\author{R.~A. Garc{\'{\i}}a}
\affiliation{Laboratoire AIM, CEA/DRF -- CNRS -- Univ. Paris Diderot -- IRFU/SAp, Centre de Saclay, F-91191 Gif-sur-Yvette Cedex, France}

\author{A. Garc{\'{\i}}a Saravia Ortiz de Montellano}
\affiliation{Max Planck Institute for Solar System Research, Justus-von-Liebig-Weg 3, D-37077 G{\"o}ttingen, Germany}
\affiliation{Stellar Astrophysics Centre, Department of Physics and Astronomy, Aarhus University, Ny Munkegade 120, DK-8000 Aarhus C, Denmark}

\author{P. B. Stetson}
\affiliation{NRC Herzberg Astronomy and Astrophysics, Dominion Astrophysical Observatory, 5071 West Saanich Road, Victoria, BC, V9E 2E7, Canada}

\author{J. De Ridder}
\affiliation{Instituut voor Sterrenkunde, Katholieke Universiteit Leuven, Celestijnenlaan 200 B, B-3001 Heverlee, Belgium}

\begin{abstract}

In cases where both components of a binary system show oscillations, asteroseismology has been proposed as a method to identify the system.
For KIC\,2568888, observed with {\it Kepler}, we detect oscillation modes for two red giants in a single power density spectrum.
Through an asteroseismic study we investigate if the stars have similar properties, which could be an indication that they are physically bound into a binary system. 
While one star lies on the red giant branch (RGB), the other, more evolved star, is either a RGB or asymptotic-giant-branch star. We found similar ages for the red giants and a mass ratio close to 1. Based on these asteroseismic results we propose KIC\,2568888 as a rare candidate binary system ($\sim 0.1\%$ chance). However, when combining the asteroseismic data with ground-based $BVI$ photometry we estimated different distances for the stars, which we cross-checked with {\it Gaia} DR2.
From {\it Gaia} we obtained for one object a distance between and broadly consistent with the distances from $BVI$ photometry. For the other object we have a negative parallax with a not yet reliable {\it Gaia} distance solution.
The derived distances challenge a binary interpretation and may either point to a triple system, which could explain the visible magnitudes, or, to a rare chance alignment ($\sim 0.05\%$ chance based on stellar magnitudes). This probability would even be smaller, if calculated for close pairs of stars with a mass ratio close to unity in addition to similar magnitudes, which may indeed indicate that a binary scenario is more favourable.

\end{abstract}


\section{Introduction} \label{sec:intro}

The {\it Kepler} space mission \citep{2010bor} obtained high-precision photometric time series data for thousands of red giants that show solar-like oscillations. In many {\it Kepler} light curves eclipses were present due to the passage or occultation of a companion star or planet. These data have led to the detection of a number of eclipsing binary systems with an oscillating red-giant component \citep[e.g.][]{2010hek,2013gau}. From theoretical predictions we expect more red giants with observable oscillation modes to belong to binary star systems \citep{2014mig}. The detection of these non-eclipsing binary systems requires different measures since they lack the distinct dips in flux in the light curves.
A new class of eccentric non-eclipsing binary systems has already been detected with {\it Kepler} data \citep[e.g.][]{2014bec,2015bec}. These binaries show ellipsoidal modulations due to strong gravitational distortions and heating that take place during periastron passage, which become visible as ``heartbeat" effects in their light curves. Their detection offered a new way of studying binary interactions as well as the evolution of such eccentric systems.

Moreover, \citet{2014mig} suggested to use asteroseismology to find potential binary systems that consist of two oscillating solar-type and/or red-giant stars. This method is applicable to high-precision long-term photometric data and independent of the inclination, separation and velocities of the binary components. 
Based on simulations of binary populations in the {\it Kepler} field of view, \citeauthor{2014mig} performed a study to predict the asteroseismic detectability of two solar-like oscillators that are gravitationally bound in a single light curve. According to their predictions there should be 200 or more so-called asteroseismic binaries detectable in {\it Kepler} long-cadence data. When considering a total number of about 200\,000 long-cadence targets, we obtain a $\sim 0.1$~per cent chance of finding an asteroseismic binary.
Additionally, the components should have a mass ratio near unity, which favors oscillations that overlap in frequency. 

So far, there are only three published cases of asteroseismic binaries that are not in eclipsing systems, and that were detected in a single {\it Kepler} light curve. \citet{2015app} reported the double-star system KIC\,7510397 (or HIP\,93511 or HD\,177412), which shows two separated oscillation-mode envelopes of two solar-like stars with typical frequencies of the oscillations at about 1\,200 and 2\,200\,$\mu$Hz. By using speckle interferometry, \citeauthor{2015app} constrained the binary orbit and determined the stellar properties of both components from asteroseismic methods. 
More recently, \citet{2017whi} presented the asteroseismic study of the binary system HD\,176465 including the detection of individual mode frequencies for two solar-like oscillating stars that are on the main sequence. The stellar oscillations of both components cover the same range in frequency from $\sim 2\,000$ to $\sim 4\,500\,\mu$Hz. Based on the derived stellar parameters, \citeauthor{2017whi} classified them as two young physically-similar solar analogues.
Furthermore, \citet{2017bec} reported the detection of an eccentric binary system that consists of a sub and a super-Nyquist oscillating red-giant star with stellar oscillations present at around $120-250\,\mu$Hz. The two binary components were found to be low-luminosity red giants of similar mass that are in the early and advanced stage of the first dredge-up event on the red-giant branch.
Another interesting system was found by \citet{2016raw}. KIC\,9246715 is a double red giant eclipsing binary, where both components have very similar masses and radii. They measured one main set of solar-like oscillations with lower amplitudes and larger oscillation mode line widths than expected, while a second set of oscillations was only marginally detectable. \citeauthor{2016raw} interpreted this as due to stellar activity and tidal forces weakening the oscillations of both stars.

These recent studies show the great potential of asteroseismology for binary systems where oscillations of the components can be detected. Here, we analyze the {\it Kepler} target KIC\,2568888, which was originally proposed to be observed in the framework of a study of the open cluster NGC\,6791. Based on asteroseismic stellar properties, {\it Gaia} parallaxes and supplementary ground- and space-based photometric measurements, we investigate if KIC\,2568888 is a rare candidate binary system with two oscillating red-giant stars.

\begin{figure}
\centering
	\includegraphics[width=0.6\columnwidth,bb= 0 0 600 600]{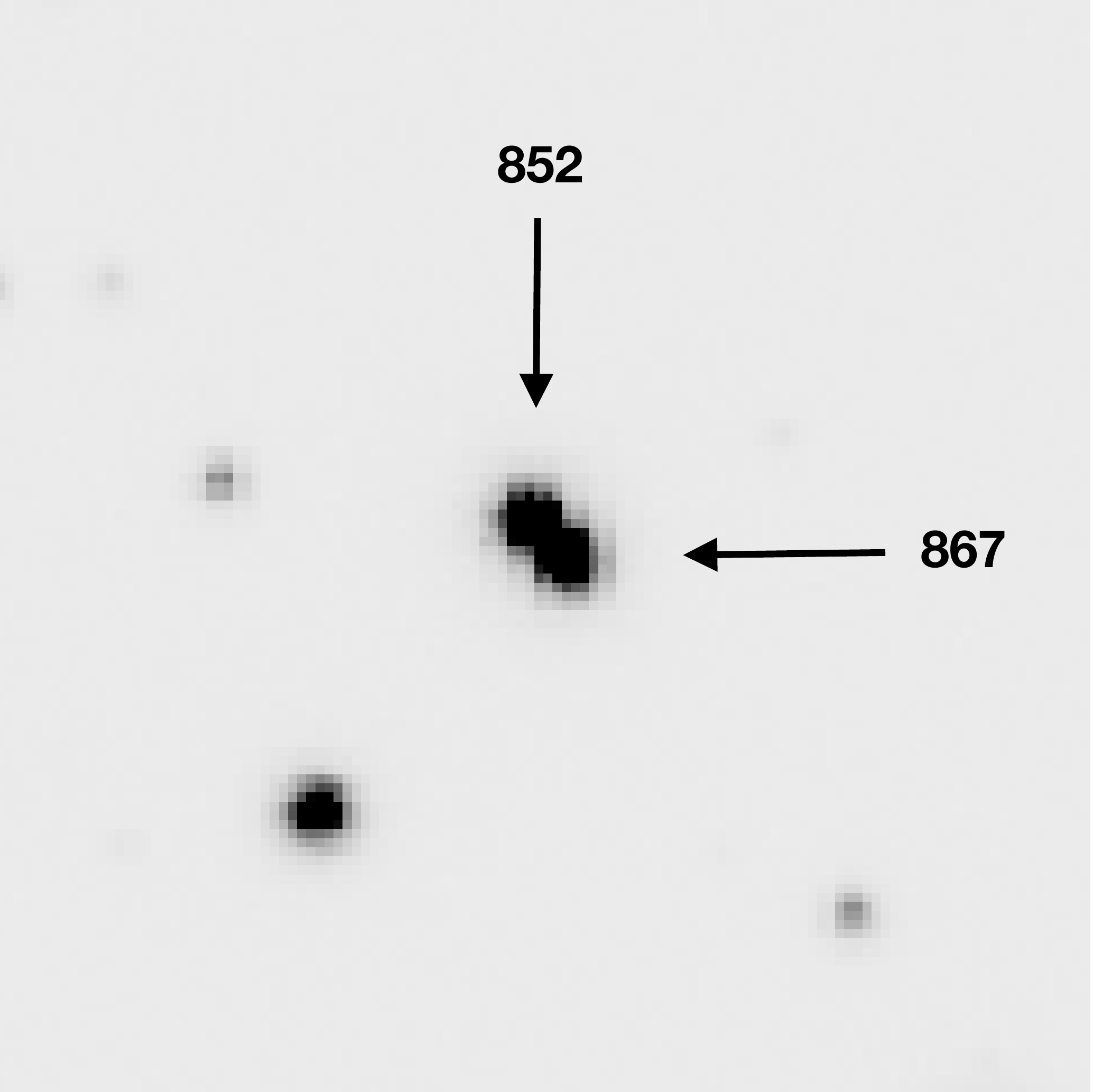}
    \caption{A $30\times 30$ arcsecond postage stamp (stack of 10 images with $0\farcs 69 < \rm seeing < 0\farcs 82$) from \citet{2003ste,2005ste} centered on our pair of red-giant stars, which have a spatial separation of $\sim1\farcs6$. The numbers are the sequential catalogue identifications for the two stars as listed in Table~\ref{tab:gaia}.}
    \label{fig:fig1}
\end{figure}

\section{Data}
\label{sec:data}

In the following, we provide an overview of the data that we used for the analysis of KIC\,2568888.

\subsection{{\it Kepler} light curve}

The basis for our asteroseismic analysis is the {\it Kepler} photometric time series of the respective stars. During each {\it Kepler} observation, pixel files were acquired for any given target star. The pixels contained within a predefined mask were then added up to create light curves \citep{2016ksci}.
When two stars are spatially coincident, as is the case here, they are observed as a single {\it Kepler} time series. To check if an optimal aperture mask was applied to derive this light curve, we inspected the individual target pixels of KIC\,2568888. We found that all pixels with stellar signal include the flux of both stars, which makes the extraction of individual light curves impossible. 
We also performed the aperture photometry with the KADACS software \citep{2011gar} with the same result. 

For the asteroseismic study we used the concatenated corrected light curve \citep{2014han} that was created from 860~days of observations during the nominal {\it Kepler} space mission with one exposure taken every $\sim 29.4$~min (long-cadence mode). From these light-curve data we determined the asteroseismic stellar parameters for both red-giant stars, which we describe in Section~\ref{sec:stellar_par}.

\subsection{APOGEE spectra}

As further constraints we took the effective temperature and metallicity that were published by the $14^{\rm th}$ data release of the Sloan Digital Sky Survey \citep[SDSS; e.g.][]{2017bla,2018abo}: $T_{\rm eff} = 4378 \pm 76$~K and $\rm [M/H]= -0.005 \pm 0.023$~dex. These atmospheric parameters were derived from APOGEE spectra by using an automated pipeline and under the assumption that KIC\,2568888 is a single stellar object with coordinates $\alpha=290^{\circ}.09354\pm0^{\circ}.00460$ and $\delta=37^{\circ}.86151\pm 0^{\circ}.00440$. 
APOGEE visited the system twice with the observations being 16 days apart. The mean measured radial velocity with respect to the barycentre is $-58.90\rm\,km\,s^{-1}$ with a radial velocity scatter of $0.18\rm\,km\,s^{-1}$.
Only the spectral lines of a single component were visible in the spectra. 
Whenever we used $T_{\rm eff}$ and [M/H] in the asteroseismic stellar parameter determination (Sec.~\ref{sec:stellar_par}), we adopted uncertainties of $\pm\, 200$~K and $\pm\, 0.3$~dex respectively. The impact of the unknown individual temperatures in particular is marginal, since red giants cover a very narrow range in $T_{\rm eff}$, which lies within the adopted uncertainties.

\subsection{Ground-based $BVI$ photometry}
\label{sec:bvi}

One of the photometric measurements in which the candidate binary is resolved is shown in Figure~\ref{fig:fig1}, where we see two stars with similar visual magnitudes in the region around KIC\,2568888. These data were obtained during the original photometric survey of the open cluster NGC\,6791 \citep{2003ste,2005ste} with additional measurements that were acquired more recently.
We analyzed the $BVI$ photometry of the two stars by using the methodology discussed in \citet{2000ste,2003ste,2005ste}. The positional and photometric measurements for both stars are given in Table~\ref{tab:gaia}. We used these magnitudes to estimate the distances in Section~\ref{sec:distances}.

\subsection{(Near-)infrared photometry}
\label{sec:nir}

In the Two Micron All Sky Survey \citep[2MASS; e.g.][]{2006skr}, the two stars are unresolved and thus observed as a single source. We report the combined $J-$band, $H-$band and $K_{\rm s}-$band magnitude in Table~\ref{tab:gaia}, which we used to investigate the distances (see Sec.~\ref{sec:distances}).

We further note that the Wide-field Infrared Survey Explorer \citep[WISE; e.g.][]{2010wri} provides combined fluxes for the pair of stars in four bands, which are all similar. 
We do not find any noticeable features such as infrared excess emission, which in case of a binary star would be circumbinary dust that is commonly observed in post-asymptotic giant branch systems \citep[e.g.][]{2013der}.
However, in case of a pole-on system, the dust disc may be outside the photometric mask that was selected for obtaining the photometry.

\subsection{{\it Gaia} DR2 parameters}
\label{sec:gaia}

\begin{figure}
\centering
	\includegraphics[width=1\columnwidth]{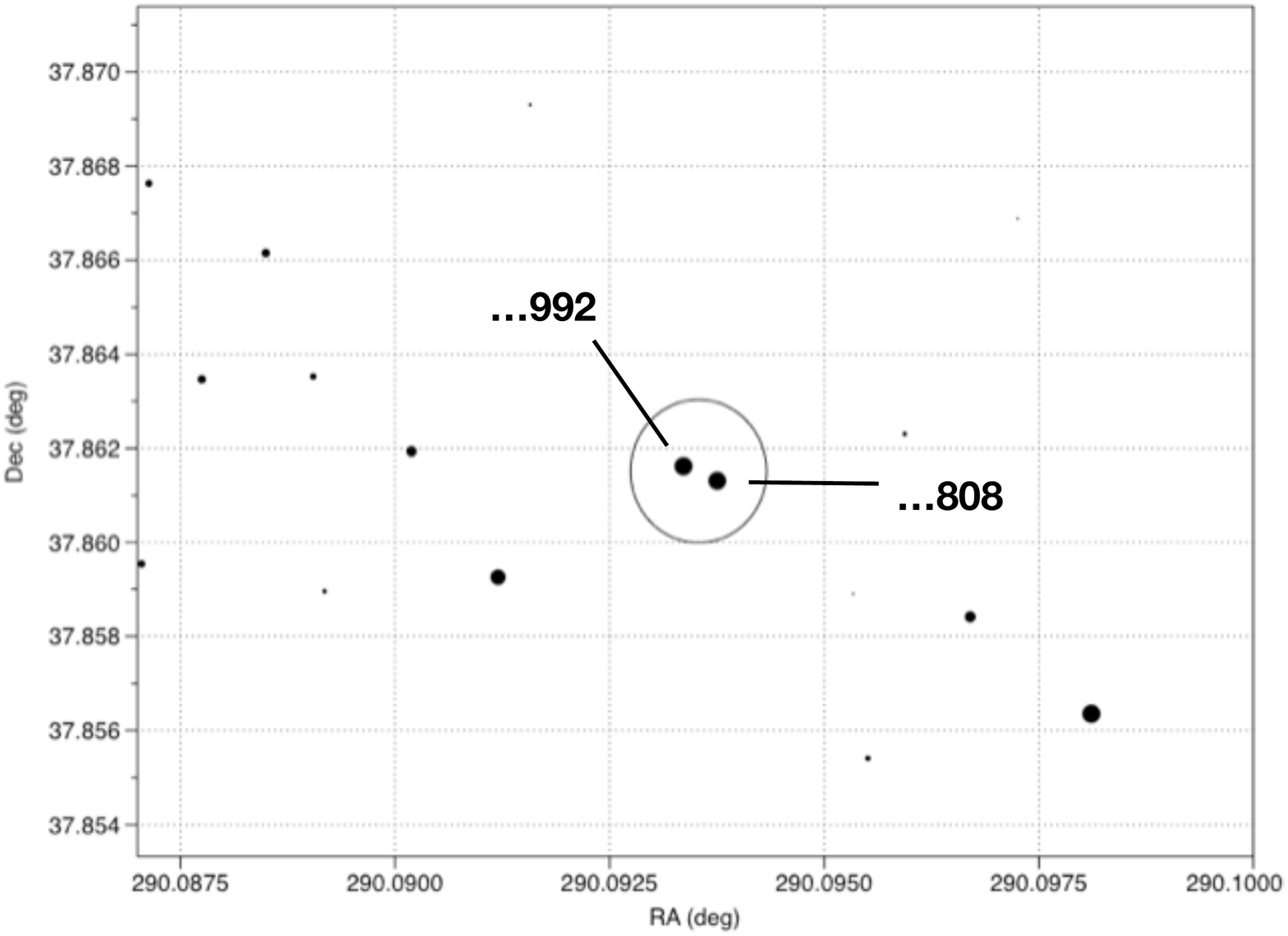}
    \caption{The position of the two stars in the {\it Gaia} view with the numbers indicating the last three digits of their source names as listed in Table~\ref{tab:gaia}.}
    \label{fig:fig1b}
\end{figure}

The second {\it Gaia} data release \citep[DR2;][]{2018gai} provided new data on the two stars that are identified as KIC\,2568888, which we report in Table~\ref{tab:gaia}. The astrometric and photometric measurements were derived from a 22 months time span of observations.
With an effective angular resolution of about $0\farcs5$, {\it Gaia} could resolve the two sources. Due to the astrometric precision of $4\cdot 10^{-5}$ arcsec we can update their angular separation to be $\sim1\farcs5765$ (Fig.~\ref{fig:fig1b}).
 
For both red giants parallaxes are provided, although for one of the stars the parallax value is negative. This is a result of the measurement process in cases where a model is fitted to noisy observations. About one quarter of all published parallaxes in {\it Gaia} DR2 are negative \citep{2018bai}. 
If a negative parallax is used to estimate the distance, it is important to treat its derivation as an inference problem by using a full Bayesian analysis, because the likelihood is not informative enough and the influence of the prior is significant. 
Following this approach, we found different distances for the two red giants with $\sim 5.6$\,kpc and $\sim 14.7$\,kpc, respectively.
It is worth noting that binarity was not taken into account in {\it Gaia} DR2. The source model that was used to derive the astrometric parameters is representative of a single star with the assumption of a uniform and rectilinear space motion relative to the solar system barycentre \citep{2018lur}. This model describes the typical helix movement for the apparent motion of a star on the sky. 
Binarity can disrupt this movement and may alter the distance estimates and proper motions of the binary components significantly.
Moreover, the pair of stars that we study is very close and they have similar visible magnitudes (see Table~\ref{tab:gaia}). This can lead to some further confusion during the analysis process, in which blending and decontamination from nearby sources were not included.

In Table~\ref{tab:gaia} we also provide the `$\rm astrometric\_excess\_noise$' ($\epsilon$), which quantifies how well the astrometric 5~$-$~parameter model fits the observations. A large value of $\epsilon$ would show that the astrometric fitting was problematic. In order to evaluate the statistical significance of this parameter, we can use the dimensionless `$\rm astrometric\_excess\_noise\_sig$' quantity ($D$). In cases where $D\leq 2$ the astrometric excess noise is considered statistically insignificant. For both stars investigated here, the provided $D$ values are larger. This indicates that the {\it Gaia} astrometric pipeline did indeed encounter some problems when fitting the astrometric model, in particular for the source 2051291674955780992, in which case a negative parallax measurement was provided.
Due to these possible complications we are not able to make any firm conclusions on the physical relation between the two stars based on the recently published {\it Gaia} data.

We further note that the resolution of the {\it Gaia} $G-$band is of the order of $\sim 0\farcs3-0\farcs5$, thus individual values for each star could be retrieved and they are of the same order of magnitude as the ground-based $V$ measurements. 
The mean red- ($G_{\rm RP}$) and blue-band ($G_{\rm BP}$) magnitudes are given for one object only. As these bands have a lower resolution of $\sim 2\farcs$, which exceeds the angular separation of our pair of stars, it is likely that the published $G_{\rm BP}$ and $G_{\rm RP}$ values represent the combined flux of both sources. 

At this moment no radial velocity data are provided by {\it Gaia} for our pair of stars.

\newcommand{\HT}{\hspace*{-0.8em}}
\begin{table*}[]
\centering
\caption{Ground-based $BVI$ photometric measurements (Sec.~\ref{sec:bvi}) and {\it Gaia} DR2 parameters (Sec.~\ref{sec:gaia}) for the two stars as well as the combined 2MASS $JHK_{\rm s}$ magnitudes (Sec.~\ref{sec:nir}). We note that the uncertainties in the {\it Gaia} distance estimates are represented by the lower and upper boundary of the 68 per cent credibility intervals.}
\label{tab:gaia}
\begin{tabular}{lrlrl}
\hline
\hline
\multicolumn{5}{c}{Ground-based $BVI$ photometry from \citet{2003ste,2005ste}}  \\ \hline
Sequential catalogue number & \multicolumn{2}{c}{867} & \multicolumn{2}{c}{852}\\
$\alpha$ (J2000.0) [deg] & \multicolumn{2}{c}{$290.09375$} & \multicolumn{2}{c}{$290.09333$}\\
$\delta$ (J2000.0) [deg] & \multicolumn{2}{c}{$+37.86144$} & \multicolumn{2}{c}{$+37.86178$}\\
$B$ [mag]  & 16.001 & \HT$\pm$ 0.004 & 16.016 & \HT$\pm$ 0.003 \\
$V$ [mag]  & 14.689 & \HT$\pm$ 0.004 & 14.621 & \HT$\pm$ 0.004\\
$I$ [mag]   & 13.258 & \HT$\pm$ 0.012 & 13.176 & \HT$\pm$ 0.006\\
\hline
\multicolumn{5}{c}{{\it Gaia} data release 2 parameters}\\ \hline
Source name & \multicolumn{2}{c}{2051291674950663808} & \multicolumn{2}{c}{2051291674955780992} \\
$\alpha$ (J2000.0) [deg] & $290.0937534437$ & \HT$\pm$ 0.0000000058 & $290.0933600817$ & \HT$\pm$ 0.0000000067\\
$\delta$ (J2000.0) [deg] & $37.8613104547$ & \HT$\pm$ 0.0000000078 & $37.8616192123$ & \HT$\pm$ 0.0000000089 \\
Parallax [mas] & 0.137 & \HT$\pm$ 0.029 & $-0.145$ & \HT$\pm$ 0.035\\
Proper motions ($\alpha^{*}$) [mas $\rm yr^{-1}$] & $-2.123$ & \HT$\pm$ 0.045 & $-3.833$ & \HT$\pm$ 0.053\\
Proper motions ($\delta$) [mas $\rm yr^{-1}$] & $-4.675$ & \HT$\pm$ 0.051 & $-7.189$ & \HT$\pm$ 0.053 \\
G [mag] & \multicolumn{2}{c}{14.25}  & \multicolumn{2}{c}{14.15} \\
G$_{\rm BP}$ [mag] & \multicolumn{2}{c}{--} & \multicolumn{2}{c}{14.70}\\
G$_{\rm RP}$ [mag] & \multicolumn{2}{c}{--} & \multicolumn{2}{c}{13.04}\\
$\epsilon$ [mas] & \multicolumn{2}{c}{0.11}  & \multicolumn{2}{c}{0.22} \\
$D$ & \multicolumn{2}{c}{2.4}  & \multicolumn{2}{c}{10.8} \\
Distance [pc] & \multicolumn{2}{c}{$5585^{+1026}_{-772}$}  & \multicolumn{2}{c}{$14737^{+3285}_{-2639}$} \\
\hline
\multicolumn{5}{c}{2MASS all-sky data release}  \\ \hline
Source name &	\multicolumn{4}{c}{2MASS J19202244+3751414}\\
$\alpha$ (J2000.0) [deg] & \multicolumn{2}{r}{$290.093536$} & \multicolumn{2}{l}{\HT$\pm$ 0.0075} \\
$\delta$ (J2000.0) [deg] & \multicolumn{2}{r}{$+37.861511$} & \multicolumn{2}{l}{\HT$\pm$ 0.0075} \\
$J$ [mag]  & \multicolumn{2}{r}{11.503}  & \multicolumn{2}{l}{ \HT$\pm$ 0.019} \\
$H$ [mag]  & \multicolumn{2}{r}{10.759} &  \multicolumn{2}{l}{ \HT$\pm$ 0.017} \\
$K_{\rm s}$ [mag]  & \multicolumn{2}{r}{10.629} & \multicolumn{2}{l}{ \HT$\pm$ 0.016}\\
\hline
\end{tabular}
\end{table*}

\section{Fourier spectrum analysis}
\label{sec:seis_par}

The typical feature of a star showing solar-like oscillations is a well-defined power excess that is visible in the Fourier power density spectrum (PDS). For KIC\,2568888 we observe excess power at $\sim 7\,\mu$Hz (star A) and $\sim 16$\,$\mu$Hz (star B), respectively, which can be attributed to two different red-giant stars. Through the analysis of individual oscillation modes we aim to get a picture of the interior structure of both stars.

\begin{figure*}
	\includegraphics[width=2.1\columnwidth]{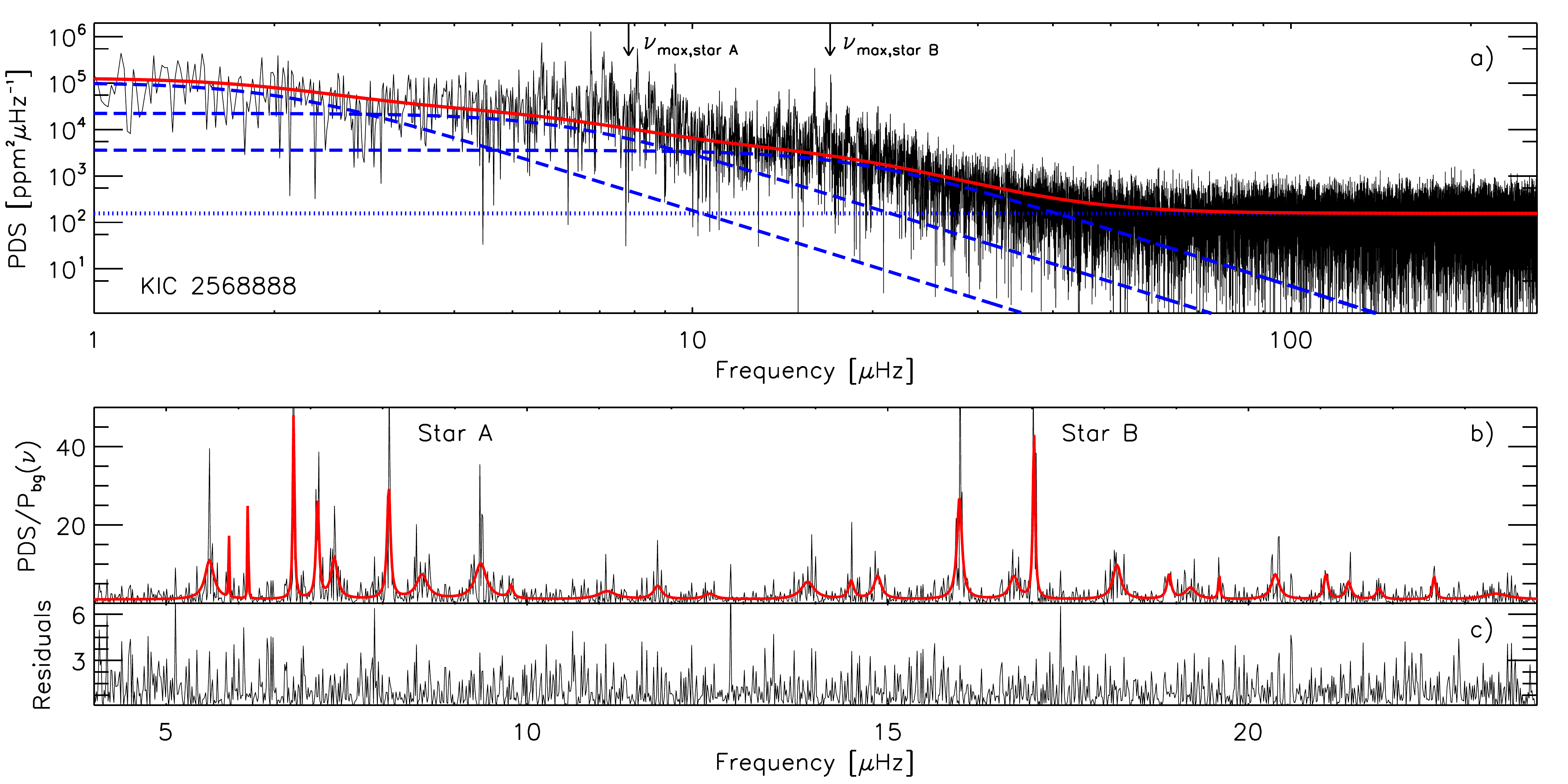}
    \caption{a) Power density spectrum (PDS; in black) of KIC\,2568888 showing three granulation components (blue dashed lines), one noise component (blue dotted line) and the global background fit (in red). The $\nu_{\rm max}$ values for stars A and B are indicated by arrows. b) Background-normalized PDS covering the oscillation modes of the two red-giant stars with the final fit to the detected oscillation modes in red and the residuals in panel c).}
    \label{fig:fig2}
\end{figure*}

\subsection{The global background model}
\label{sec:bg}

The oscillation modes are superimposed on a granulation background, which we need to define before studying the oscillations. We chose a model similar to the descriptions used by \citet{2014kal} with the contribution of three granulation background components:
\begin{equation}
    P_{\rm bg}(\nu)=n_{\rm wh}+\eta(\nu)^2 \left(\sum^{3}_{\rm i=1}\frac{A_{\rm i}}{1+(\nu/b_{\rm i})^4}\right).
	\label{eq:p_bg}
\end{equation}
Here, $n_{\rm wh}$ represents the white noise. Each granulation component is defined by a Lorentzian-like function with a characteristic amplitude $A_{\rm i}$, frequency $b_{\rm i}$ and a fixed exponent of 4. The stellar granulation is further influenced by an attenuation $\eta$, which arises due to discrete time sampling of the flux measurements.

As a parameter estimation method we employed a Bayesian Markov Chain Monte Carlo (MCMC) framework with affine-invariant ensemble sampling, as implemented in the {\sc Emcee}\footnote{{\sc Emcee}: The MCMC Hammer, \url{http://dfm.io/emcee/}} routine \citep{2013for}, to explore the parameter space of the granulation background components (eq.~\ref{eq:p_bg}).
After convergence, we used the MCMC chains to estimate the posterior probability density functions for each parameter. We adopted the medians of these distributions as an estimate of the expectation values for the parameters and their 16th and 84th percentiles as standard uncertainties.
In panel a) of Figure~\ref{fig:fig2} we show the PDS of KIC\,2568888 and the global background fit. 
The oscillation regions of the two red-giant stars are marginally overlapping at the edges. The parameter $\nu_{\rm max}$ represents the frequency of maximum oscillation power, which we define as the center of the oscillation power envelope.
In order to determine $\nu_{\rm max}$ for each red giant, we first corrected the PDS by the global background and then we fitted a model with two Gaussian functions to the normalized PDS.
The global background parameters and $\nu_{\rm max}$ values are listed in Table~\ref{tab:bg}.
We further note that the amplitudes of the power excesses are on the lower edge of the empirical $\nu_{\rm max}$--amplitude relation \citep[e.g.][]{2011kje,2011ste,2011hub,2012bmos}. Although that does not provide decisive information about our pair of stars, it shows that the presence of two stars provides a `diluted' light curve, which results in the observation of decreased amplitudes of the oscillations.

\subsection{Oscillations}
\label{sec:dnu}

\begin{figure}
	\includegraphics[width=1.\columnwidth]{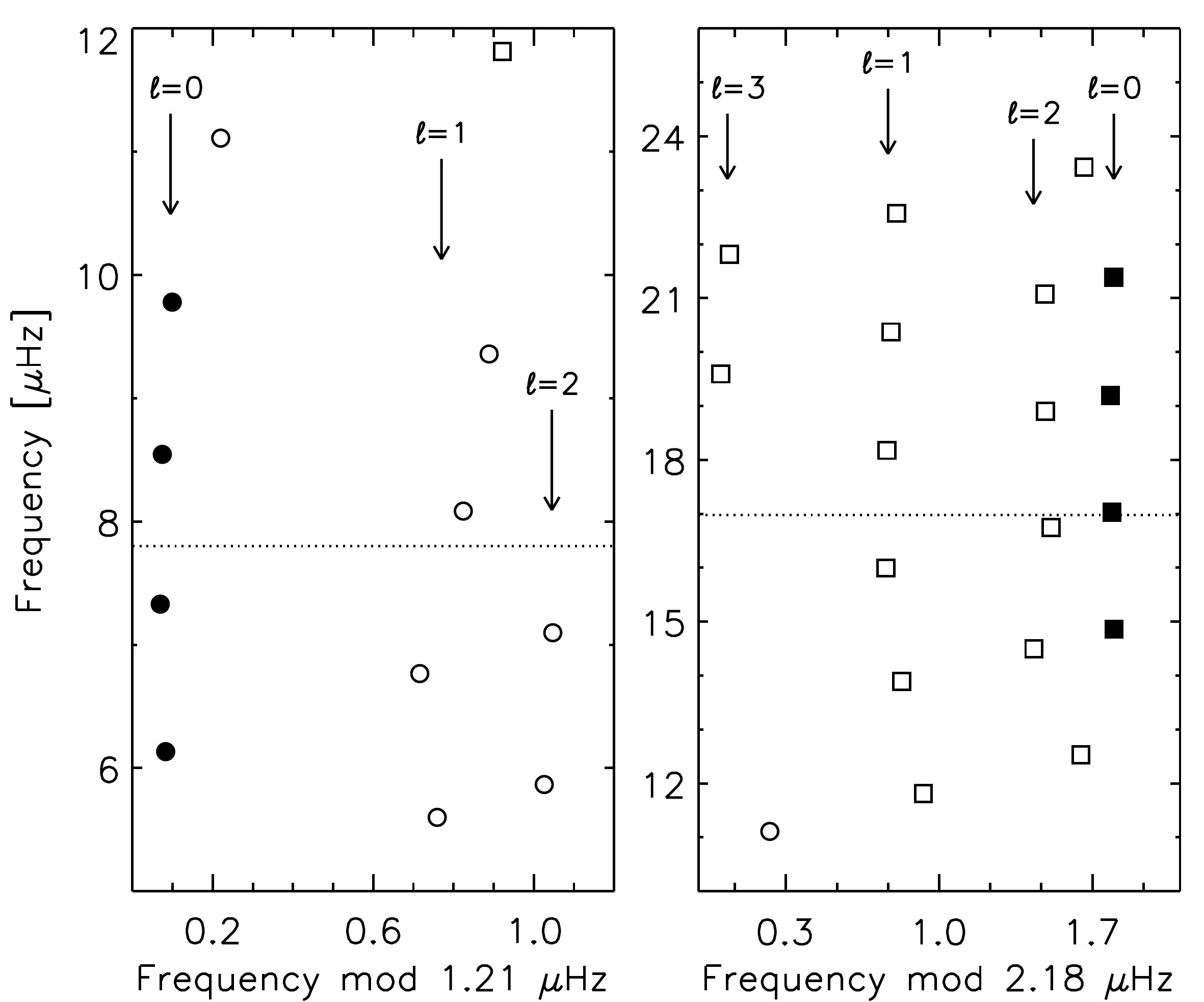}
    \caption{{\'E}chelle diagrams for stars A (left) and B (right) with extracted mode frequencies (A: circles; B: squares) that form vertical ridges corresponding to different spherical degrees $\ell$. Filled symbols show frequencies that were used to determine $\Delta\nu$ and horizontal dotted lines represent $\nu_{\rm max}$ values.}
    \label{fig:fig3}
\end{figure}

Another asteroseismic parameter of interest is the mean large frequency separation $\Delta\nu$, i.e. the frequency spacing between pressure (p) modes of the same spherical degree $\ell$ and consecutive radial order $n$. Here, we used the continuous wavelet transform-based peak detection method developed by \citet{2018gar} to search for all significant Lorentzian-like peaks in the background-normalized PDS.
In the current analysis, we applied this automated peak detection algorithm with a signal-to-noise threshold of 1.5. 
As a measure of the statistical significance of each peak we compared the Akaike Information Criterion \citep[AIC;][]{1998aka} of a PDS model including the peak and a model without it. The AIC difference between the two models is similar to a log-likelihood difference with a penalisation for the number of degrees of freedom. The model with a lower AIC value is preferred. For more details about the peak detection we refer the interested reader to \citet{2018gar}. 
In addition to the frequencies of the peaks, the algorithm provided initial values for their amplitudes and line widths.
Based on these estimates we used a maximum-likelihood method (MLE) to optimize all variable parameters simultaneously. From this final MLE fit we estimated the values for the frequencies, amplitudes and line widths of the oscillation modes as well as their uncertainties, which we report in Table~\ref{tab:pb}. We note that the mode amplitudes are given in units of the background-normalized power density spectrum.
In panel b) of Figure~\ref{fig:fig2} we present the background-normalized PDS including the model fit. The residuals in panel c) show that only noise is left in the PDS after the fit.
In the following, we assigned the spherical degree and the acoustic radial order to the set of detected frequencies by using the asymptotic relation \citep{1980tas}. For both stars in KIC\,2568888 we detected several radial orders of $\ell=0,1$ and 2 modes and two $\ell=3$ modes (Table~\ref{tab:pb}) that are visible as vertical ridges in the \'{e}chelle diagrams \citep{1983gre} in Figure~\ref{fig:fig3}. As a further note, no clear evidence of mixed modes were present in the power density spectrum. Therefore, no evolutionary stage determination based on period spacings was possible for either red-giant star \citep[e.g.][]{2011mos,2013ste}. A preliminary study by \citet{2017the} presented that about 50\% of red giants in detached eclipsing binaries show only p-dominated mixed modes compared to about 4\% of red giants not known to be in binary systems. This may hint to a binary scenario for KIC\,2568888.

For each red giant, we computed the mean large frequency spacing $\Delta\nu$ from a linear fit through the set of four central $\ell=0$ modes (marked with asterisks in Table~\ref{tab:pb} and with filled symbols in Fig.~\ref{fig:fig3}) that were unambiguously assigned to the respective star. According to the asymptotic relation for $\ell=0$ modes, the slope parameter of each fit represents $\Delta\nu$ and the intercept is related to the phase term $\epsilon$. We report the mean large frequency spacings for stars A and B in Table~\ref{tab:bg}.
In addition, we derived local values ($\Delta\nu_{\rm c}$, $\epsilon_{\rm c}$) as these are proposed to provide information about the evolutionary stage of red giants \citep{2012kal}.
Based on $\Delta\nu_{\rm c, star\,B}=2.168\pm0.021\,\mu$Hz and $\epsilon_{\rm c, star\,B}=0.85\pm0.01$, star B is a red-giant-branch (RGB) star. Star A is a more evolved red giant which may be in the asymptotic giant phase (AGB) of stellar evolution with measured $\Delta\nu_{\rm c, star\,A}=1.202\pm0.012\,\mu$Hz and $\epsilon_{\rm c, star\,A}=0.10\pm0.02$.

\begin{table}
	\centering
	\caption{Global background, asteroseismic and stellar parameters from scaling relations (sr) for both stars present in the light curve of KIC\,2568888.}
	\label{tab:bg}
	\begin{tabular}{lrlrl}
		\hline
		\hline
		Parameter & \multicolumn{2}{c}{Star A} & \multicolumn{2}{c}{Star B}\\
		\hline
		$n_{\rm wh}$ [ppm$^2\mu$Hz$^{-1}$] & \multicolumn{2}{r}{$156$} &  \multicolumn{2}{l}{\HT $^{+1}_{-1}$}\\
		$A_1$ [$\rm ppm^2\mu Hz^{-1}$] & \multicolumn{2}{r}{$103\,617$} &  \multicolumn{2}{l}{\HT $^{+16\,732}_{-15\,981}$}\\
		$b_1$ [$\mu$Hz] & \multicolumn{2}{r}{$2.0$} &  \multicolumn{2}{l}{\HT $^{+0.3}_{-0.2}$}\\
		$A_2$ [$\rm ppm^2\mu Hz^{-1}$] & \multicolumn{2}{r}{$22\,301$} &  \multicolumn{2}{l}{\HT $^{+13\,405}_{-12\,446}$}\\
		$b_2$ [$\mu$Hz] & \multicolumn{2}{r}{$6.2$} &  \multicolumn{2}{l}{\HT $^{+1.9}_{-2.6}$}\\
		$A_3$ [$\rm ppm^2\mu Hz^{-1}$] & \multicolumn{2}{r}{$3\,590$} &  \multicolumn{2}{l}{\HT $^{+1\,684}_{-1\,235}$}\\
		$b_3$ [$\mu$Hz] & \multicolumn{2}{r}{$18.8$} &  \multicolumn{2}{l}{\HT $^{+1.9}_{-1.7}$}\\
		\hline
$\nu_{\rm max}$ [$\mu$Hz] & $7.82$ & \HT $\pm 0.23$ & $16.98$ & \HT $\pm 0.41$\\
		$\Delta\nu$ [$\mu$Hz] & $1.210$ & \HT $\pm 0.008$ & $2.177$  & \HT $\pm 0.011$\\
		Evolutionary state & \multicolumn{2}{c}{AGB/RGB} & \multicolumn{2}{c}{RGB}\\
		\hline
		$M_{\rm sr}$ [$\rm M_{\sun}$] & $1.36$ & \HT $\pm 0.09$ & $1.32$ & \HT $\pm 0.09$\\
		$R_{\rm sr}$ [$\rm R_{\sun}$] & $25.11$ & \HT $\pm 0.68$ & $16.84$ & \HT $\pm 0.45$\\
		$\bar{\rho}_{\rm sr}$ [$\rm \bar{\rho}_{\sun}\times10^{-3}$] & $0.086$ & \HT $\pm 0.002$ & $0.277$ & \HT $\pm 0.005$\\
		$\log g_{\rm sr}$ (cgs) & $1.770$ & \HT $\pm 0.008$ & $2.107$ & \HT $\pm 0.009$\\
		\hline
	\end{tabular}
\end{table}

\section{Determination of stellar parameters}
\label{sec:stellar_par}

We measured the global seismic parameters ($\nu_{\rm max}$ and $\Delta\nu$) to derive the stellar properties of stars A and B by using different asteroseismic methods.
Here, we use the stellar parameters to investigate if the two red giants under study could potentially be physically bound in a binary system.

\subsection{Asteroseismic scaling relations}

One method to determine the stellar parameters of red giants is based on the asteroseismic scaling relations \citep[SR;][]{1986ulr,1991bro,1995kje}. These equations require reference values often taken from the Sun and thus it is implicitly assumed that the internal stellar structure is similar for all stars of different masses, metallicities and evolutionary stages. From observations and theoretical predictions we know that this is not the case. Different studies pointed out discrepancies in the derived asteroseismic stellar parameters of red-giant stars \citep[e.g.][]{2010hub,2016gau,2018the} even though several modifications to the scaling relations were proposed in order to improve the precision of these parameter estimates \citep[e.g.][]{2011whi,2012mig,2013mos,2013hek,2016gug,2016sha,2017gug,2017rod,2017via,2018the}.

In the current study, we employ empirical reference values ($\nu_{\rm max,ref}=3137\pm 45\,\mu$Hz, $\Delta\nu_{\rm ref}=130.8\pm 0.9\,\mu$Hz) that were derived from a combined asteroseismic and binary analysis of three RGB stars \citep{2018the}. By using these reference values the metallicity, temperature, mass dependence of stars as well as surface effects are incorporated in the SR.
Based on the global seismic parameters, the spectroscopic effective temperature and metallicity from APOGEE spectra, and the empirical reference values we computed the asteroseismic stellar parameters for both stars. 
We note that the formal uncertainties in the derived stellar parameters are larger due to our adopted uncertainty of $\pm 200$\,K in temperature since we lack individual $T_{\rm eff}$ values for stars A and B. 
The stellar parameters are reported in Table~\ref{tab:bg}.

\begin{table}
	\centering
	\caption{Stellar parameters from grid-based modeling (gbm) by using PARSEC isochrones for both stars. We note that for star A we also find a matching RGB model which is less likely than the AGB model.}
	\label{tab:gbm}
	\begin{tabular}{lrlrlrl}
		\hline
		\hline
		Parameter & \multicolumn{4}{c}{Star A} & \multicolumn{2}{c}{Star B}\\
		\hline
		$M_{\rm gbm}$ [$\rm M_{\sun}$] & $1.35$ & \HT $\pm 0.20$ & $1.37$ & \HT $\pm 0.18$ & $1.37$ & \HT $\pm 0.21$\\
		$R_{\rm gbm}$ [$\rm R_{\sun}$] & $25.07$ & \HT $\pm 1.26$ & $25.20$ & \HT $\pm 1.24$ & $17.05$ & \HT $\pm 1.01$\\
		$\bar{\rho}_{\rm gbm}$ [$\rm \bar{\rho}_{\sun}\times10^{-3}$] & $0.086$ & \HT $\pm 0.004$ & $0.086$ & \HT $\pm 0.004$ &  $0.277$ & \HT $\pm 0.010$\\
		$\log g_{\rm gbm}$ (cgs) & $1.770$ & \HT $\pm 0.019$ & $1.770$ & \HT $\pm 0.019$ & $2.112$ & \HT $\pm 0.026$\\
		$\log (L/ \rm L_{\sun})_{\rm gbm}$ & $2.30$ & \HT $\pm 0.12$ & $2.29$ & \HT $\pm 0.10$ & $1.99$ & \HT $\pm 0.11$\\
		$T_{\rm eff, gbm}$ [K] & $4324$ & \HT $\pm 188$ & $4301$ & \HT $\pm 161$ & $4419$ & \HT $\pm 182$\\
		age$_{\rm gbm}$ [Gyr] & $3.6$ & \HT $\pm 1.5$ & $3.2$ & \HT $\pm 1.2$ & $3.7$ & \HT $\pm 1.7$\\
		Evolutionary state & \multicolumn{2}{c}{AGB} & \multicolumn{2}{c}{RGB} & \multicolumn{2}{c}{RGB}\\ 
		\hline
	\end{tabular}
\end{table}

\subsection{Grid-based modeling}
\label{sec:gbm}

In addition to the determination of stellar parameters through SR one can also use a precomputed grid of stellar isochrones to find the best-fit model to the observational data. For our grid-based modeling \citep[GBM;][]{2011gai} approach we computed a set of stellar isochrones with the PAdova and TRieste Stellar Evolution Code \citep[PARSEC;][]{2012bres}. These isochrones extend from the lower main sequence up to the asymptotic giant branch for stars between $0.1\,\rm M_{\sun}$ and $ 12\,\rm M_{\sun}$ with ages ranging from $\sim 4$\,Myr to 13.2\,Gyr and metallicities in the range $0.0005\le \rm Z \le 0.07$ (corresponding to $-1.49\le \rm [M/H] \le +0.78$). For low mass stars, mass loss due to stellar winds is incorporated during the RGB phase according to the empirical formula by \citet{1975rei} with an efficiency factor of 0.2. We obtained this grid of stellar models through the CMD web interface at OAPD\footnote{\url{http://stev.oapd.inaf.it/cmd/}}.

The stellar parameters were extracted from this grid using an independent implementation of the likelihood method described by \citet{2010bas}, where the likelihood of each model was computed from a given set of observed parameters. In this case, we used $\nu_{\rm max}$ and $\Delta\nu$ from the asteroseismic analysis and atmospheric parameters ($T_{\rm eff},\rm [M/H]$) provided by APOGEE to search for matching stellar models. 
For the computation of $\Delta\nu$ and $\nu_{\rm max}$ for the models, we employed the scaling relations with the empirical reference values as stated in the previous section.
Based on a Monte Carlo method, we obtained the stellar parameters including their uncertainties for each star from the center and width of a Gaussian fit through the total likelihood distribution of 1\,000 perturbations, which we report in Table~\ref{tab:gbm}. 
We note that the uncertainties of the GBM results are larger due to the lack of individual $T_{\rm eff}$ and [M/H] measurements for our pair of stars.

In addition to individual ages of the two red-giant stars, the GBM approach provides an indication of which evolutionary state is favored \citep{2017bhek}. For star A we found two solutions in different evolutionary stages that matched the observations, a red-giant-branch and a more evolved asymptotic-giant-branch model (Table~\ref{tab:gbm}). Based on our optimization method, we obtained a marginally higher statistical significance for the solution on the asymptotic giant branch.
For both stars, we found the same evolutionary stages from stellar models and from the study of the local phase terms (Sec.~\ref{sec:dnu}). To check the results from PARSEC, we repeated the GBM analysis with stellar isochrones from the BaSTI\footnote{\url{http://albione.oa-teramo.inaf.it/}} \citep{2004pie} code and obtained consistent results.

\subsection{UniDAM}
\label{sec:distances}

UniDAM \citep{2017min,2018min} is a Bayesian isochrone fitting tool that can use different combinations of measured physical parameters (e.g. $R, T_{\rm eff}, \log g, \rm [M/H]$) as well as {\it Gaia} parallaxes as inputs to determine stellar masses, ages and distances. We used this tool for a further test to constrain the age and the distance of each star independently by comparing our final $R$ and $\log g$ values from SR (Table~\ref{tab:bg}) together with the $BVI$ photometry (Table~\ref{tab:gaia}) with PARSEC models (same set of isochrones as described in Section~\ref{sec:gbm}).
Since it is not known which photometric component corresponds to which asteroseismic signal (stars A and B), we employed both possible combinations of magnitudes with rather similar results. 
As a reference, we chose the result with the better fit to the photometry with a $\chi^2$ probability close to 1. Based on this approach, we derived age estimates for both stars that were consistent with those derived from the asteroseismic analysis, while the apparent distance moduli turned out to be different with $\mu_{\rm d,A} = 14^m.63 \pm 0.12$ and $\mu_{\rm d,B} = 13^m.96 \pm 0.15$.

\begin{figure}
	\includegraphics[width=1\columnwidth]{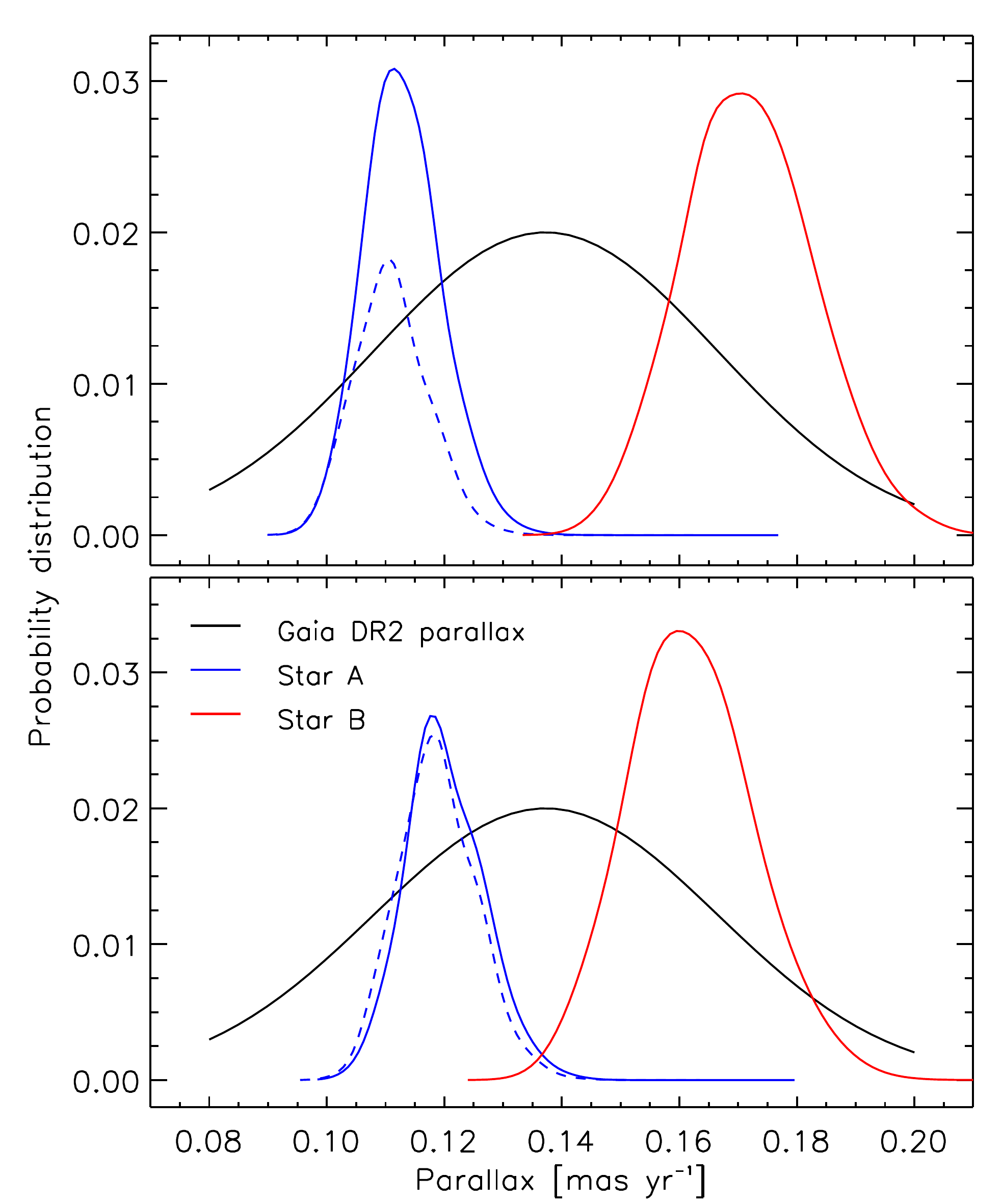}
    \caption{Parallax probability distribution functions (PDF) for the observed parallax measurement for source 2051291674950663808 from {\it Gaia} DR2 (in black) and asteroseismic stars A (in blue) and B (in red) as derived from UniDAM (Sec.~\ref{sec:distances}). For star A we show the PDFs for models in both evolutionary stages (RGB: solid, AGB: dashed). Upper and lower panels represent different combinations of asteroseismic $R$ and $\log g$ values and ground-based $BVI$ photometry, since we are not able to match the asteroseismic components with their photometric counterparts. In both cases the results are similar.}
    \label{fig:fig4}
\end{figure}

In Figure~\ref{fig:fig4} we show the {\it Gaia} DR2 parallax probability distribution function (PDF) for the star with the positive {\it Gaia} parallax measurement and the parallax PDFs as derived by UniDAM from $BVI$ photometry for both stars. From UniDAM, we computed a lower parallax value for star A and a higher parallax value for star B, while the positive {\it Gaia} parallax value lies in between with its uncertainties covering the individual PDFs of both stars.

The discrepancies in the distance estimates is not surprising due to the fact that the apparent magnitudes are very similar, while we detected two power excesses with different $\nu_{\rm max}$ values and thus expect the two stars to have different radii (see Tables~\ref{tab:bg} and~\ref{tab:gbm}).
With similar effective temperatures, which can be assumed given the similar observed colors, different radii should lead to different absolute luminosities and thus absolute magnitudes. In our analysis, this difference is on the order of $0^m.65$ which can only partly be explained by the uncertainties in the models, the photometric calibration and the extinction model that were used.

In addition, we can test if we find solutions for the two stars assuming that they are in a binary system. In this case, we took as a constraint the combined apparent magnitude for KIC\,2568888 from 2MASS (Table~\ref{tab:gaia}), where the pair of stars could not be spatially resolved. The combined magnitude should then match the predicted magnitudes from the best-fitting models for both stars. 
We selected a pair of models such that the following conditions are fulfilled: (1) Metallicity and age (as computed from GBM) are the same within uncertainties for both models; (2) $R$ and $\log g$ are within $4\,\sigma$ uncertainties from values derived from the asteroseismic analysis; and (3) $BVI$ photometry for each model and combined $JHKs$ photometry match the observed values.
Based on these criteria, we found solutions that give an age of $\sim 3.4$\,Gyr, a metallicity of $\sim -0.06$\,dex and a distance modulus of $\sim 14^m.4$ or $\sim 7.5$\,kpc. We note however that the $\chi^2$ probability for this model pair is rather low on the order of $10^{-9}$.

\section{Discussion and conclusions}

In the {\it Kepler} light curve of KIC\,2568888 we detected asteroseismic signals of two red giants. The asteroseismic analysis leads us to the following conclusions:
\begin{itemize}
\item[-] The similar ages ($\sim 3.6$ and $\sim 3.7$\,Gyr) for stars A and B, and a mass ratio close to unity support a possible binary scenario, where KIC\,2568888 is comprised of either two RGB components or a RGB/AGB combination.
\item[-] If KIC\,2568888 is indeed an asteroseismic binary system then this would be a very interesting and rare candidate binary according to \citet{2014mig}, who pointed out that the detection of a binary system in this configuration is possible, yet not that common. According to their study, the overall probability of detecting two solar-like oscillating binaries in a single {\it Kepler} time series is of the order of $0.1$~per cent.
\item[-] If the derived ages of stars A and B are accurate, then it is unlikely that the two stars belong to the old ($\sim 8$\,Gyr) open cluster NGC\,6791 \citep[e.g.][]{2018mar} as a member of which KIC\,2568888 was proposed for observations. Moreover, the provided APOGEE radial velocity for the system (about $-58\rm\,km\,s^{-1}$) is not in line with either the radial velocity of the cluster \citep[$-47.40\pm0.13\rm\,km\,s^{-1}$;][]{2014tof} or published APOGEE radial velocities of known red-giant cluster members ranging from about $-43$ to about $-50\rm\,km\,s^{-1}$.
\end{itemize}
In addition, we computed distance estimates for the two stars, which are challenging a binary interpretation. From ground-based $BVI$ photometry that we combined with the asteroseismic radii and logarithmic surface gravities, we obtained different distance moduli for both stars with $\mu_{\rm d,A} = 14^m.63 \pm 0.12$ and $\mu_{\rm d,B} = 13^m.96 \pm 0.15$, respectively. Comparing this with recently published {\it Gaia} data, we found a consistent distance estimate for the star with the positive {\it Gaia} parallax measurement, while for the other star a negative parallax is provided that results in a different distance measurement. Even though this distance may be correct, we do not consider it reliable due to the strong influence of the chosen prior in the Bayesian analysis as well as the {\it Gaia} `$\rm astrometric\_excess\_noise$' flag, which indicates that the astrometric fitting of this source in particular was ambiguous. This leads us to the conclusion that:

\begin{itemize}
\item[-] If the discrepancy in the distances is true, then this could indicate that the pair of stars is not gravitationally bound and a chance alignment.
\end{itemize}

To calculate the probability ($p_{\rm chance}$) of such a close pair happening as a chance alignment, we selected all stars in the observed region of NGC\,6791 with $m_{\rm B} > 16^m.1$, which corresponds to the apparent magnitude of the fainter of the two components. We found 164 stars that are spread over $0.07$ square degrees.
The probability of having a chance companion for a star at an angular separation of $s \approx 1\farcs6 \approx 0^{\circ}.00044$, as derived from the coordinates of the two components of KIC\,2568888, is given by $p_{\rm chance} = s^2 \rho$, where $\rho$ is the number density of stars on the sky. 
We have $\rho = 164 / 0.07 \approx 2\,300$ stars per square degree, thus we calculated $p_{\rm chance}$ to be $4.6 \times 10^{-4}$ or about 0.05 per cent. This value marks the upper limit of the chance alignment probability, which would decrease further if we use a mass ratio close to 1 as an additional constraint for the pairs of stars that are considered in the calculation.
We also note that we obtained a similar result when computing $\rho$ from 2MASS stars in the same area. This brings us to the final conclusions that:
\begin{itemize}
\item[-] If the stars of KIC\,2568888 are not components of a binary system, then it would be a very rare case of an optical double system with $\sim 0.05$\% chance based on the observed magnitudes.
\item[-] If the stellar radii of stars A and B are accurate and these stars are gravitationally bound, then there might be a third star to account for the excess flux. Additional radial velocity measurements from APOGEE and {\it Gaia} could potentially provide a test if either of the two stars is itself a binary. This would explain the visual magnitudes of the two observed oscillating red-giant stars, and thus the discrepancies in the distance estimates.
\end{itemize}

In any case, it will be interesting to see the {\it Gaia} end of mission products for this pair of stars. With a more complete set of astrometric and photometric parameters at hand, e.g. reliable proper motions and parallaxes for each component, the {\it Gaia} final data release may provide the only possibility to solve this issue in the near future. Since binarity will be taken into account in the prospective {\it Gaia} data analysis, we propose KIC\,2568888 as a strong candidate for further binary investigation.

\acknowledgments

We especially thank Y. Elsworth for many useful comments. Moreover, we are very grateful for the valuable suggestions and comments from the referee. This work was funded by the European Research Council under the European Community's Seventh Framework Programme (FP7/2007-2013) / ERC grant agreement no. 338251 (StellarAges). RAG acknowledges the support from the CNES. JDR gratefully acknowledges the support from the Belgian Federal Science Policy Office (Belspo, Gaia-DPAC). This research was undertaken in the context of the International Max Planck Research School for Solar System Science at the University of G{\"o}ttingen. This work has made use of data from the European Space Agency (ESA) mission {\it Gaia} (\url{https://www.cosmos.esa.int/gaia}), processed by the {\it Gaia} Data Processing and Analysis Consortium (DPAC, \url{https://www.cosmos.esa.int/web/gaia/dpac/consortium}). Funding for the DPAC has been provided by national institutions, in particular the institutions participating in the {\it Gaia} Multilateral Agreement. This publication makes use of data products from the Two Micron All Sky Survey, which is a joint project of the University of Massachusetts and the Infrared Processing and Analysis Center/California Institute of Technology, funded by the National Aeronautics and Space Administration and the National Science Foundation.

\appendix

\section{List of detected frequencies for KIC\,2568888}

We provide a list of extracted frequencies of oscillation modes including their amplitudes and line widths in Table~\ref{tab:pb}.

\begin{table}
\centering
\caption{Extracted frequencies, amplitudes and line widths of modes that were fitted with Lorentzian (resolved peaks) or $\rm sinc$ functions (unresolved peaks). In the latter case, the line width is not given. Modes marked with asterisks were used to determine $\Delta\nu$ (Sec.~\ref{sec:dnu}).
We indicate more than one spherical degree $\ell$ and radial order $n$ in cases where two modes are overlapping such that they could not be fitted individually.}
\label{tab:pb}
\begin{tabular}{cclrccr}
\hline
\hline
Star & $n$ & $\ell$ & Frequency [$\mu$Hz] & Amplitude [a.u.] & Line width [$\mu$Hz]& AIC\\
\hline
A&    5 &       1 & 5.60 $\pm$ 0.02 & 1.53 $\pm$ 0.23 & 0.08 $\pm$ 0.06 & 158.56 \\
 &     5 &       2 & 5.87 $\pm$ 0.01 & 0.99 $\pm$ 0.48 & - & 6.09 \\
 &     6 &       0* & 6.13 $\pm$ 0.01 & 1.30 $\pm$ 0.53 & - & 22.66 \\
 &     6 &       1 & 6.77 $\pm$ 0.01 & 1.43 $\pm$ 0.31 & 0.01 $\pm$ 0.01 & 162.13 \\
 &     6 &       2 & 7.10 $\pm$ 0.01 & 1.39 $\pm$ 0.25 & 0.02 $\pm$ 0.02 & 90.58 \\
 &     7 &       0* & 7.33 $\pm$ 0.02 & 1.31 $\pm$ 0.20 & 0.05 $\pm$ 0.03 & 78.16 \\
 &     7 &       1 & 8.08 $\pm$ 0.01 & 1.59 $\pm$ 0.25 & 0.03 $\pm$ 0.02 & 162.95 \\
 &     8 &       0* & 8.54 $\pm$ 0.03 & 1.35 $\pm$ 0.16 & 0.09 $\pm$ 0.04 & 75.73 \\
 &     8 &       1/2 & 9.36 $\pm$ 0.02 & 1.65 $\pm$ 0.17 & 0.10 $\pm$ 0.03 & 156.83 \\
 &     9 &       0* & 9.78 $\pm$ 0.03 & 0.57 $\pm$ 0.20 & 0.03 $\pm$ 0.03 & 10.14 \\
 &     9/10 &       2/0 & 11.11 $\pm$ 0.06 & 0.97 $\pm$ 0.17 & 0.15 $\pm$ 0.08 & 22.81 \\
 \hline
B&    5 &       1 & 11.81 $\pm$ 0.03 & 0.86 $\pm$ 0.15 & 0.07 $\pm$ 0.05 & 21.16 \\
 &     5/6 &       2/0 & 12.53 $\pm$ 0.06 & 0.63 $\pm$ 0.18 & 0.09 $\pm$ 0.07 & 2.71 \\
 &     6 &       1 & 13.89 $\pm$ 0.03 & 1.24 $\pm$ 0.15 & 0.11 $\pm$ 0.05 & 71.53 \\
 &     6 &       2 & 14.49 $\pm$ 0.02 & 0.79 $\pm$ 0.16 & 0.04 $\pm$ 0.03 & 15.35 \\
 &     7 &       0* & 14.86 $\pm$ 0.02 & 1.07 $\pm$ 0.16 & 0.06 $\pm$ 0.03 & 51.34 \\
 &     7 &       1 & 15.99 $\pm$ 0.01 & 1.83 $\pm$ 0.23 & 0.04 $\pm$ 0.02 & 281.66 \\
 &     7 &       2 & 16.75 $\pm$ 0.03 & 1.13 $\pm$ 0.17 & 0.07 $\pm$ 0.03 & 50.88 \\
 &     8 &       0* & 17.03 $\pm$ 0.01 & 1.55 $\pm$ 0.29 & 0.02 $\pm$ 0.01 & 157.77 \\
 &     8 &       1 & 18.18 $\pm$ 0.02 & 1.37 $\pm$ 0.16 & 0.07 $\pm$ 0.02 & 118.12 \\
 &     8 &       2 & 18.90 $\pm$ 0.02 & 0.83 $\pm$ 0.16 & 0.04 $\pm$ 0.02 & 21.82 \\
 &     9 &       0* & 19.20 $\pm$ 0.04 & 0.89 $\pm$ 0.15 & 0.09 $\pm$ 0.04 & 22.34 \\
 &     8 &       3 & 19.59 $\pm$ 0.01 & 0.52 $\pm$ 0.17 & 0.01 $\pm$ 0.01 & 9.87 \\
 &     9 &       1 & 20.37 $\pm$ 0.02 & 1.15 $\pm$ 0.15 & 0.07 $\pm$ 0.03 & 74.80 \\
 &     9 &       2 & 21.07 $\pm$ 0.02 & 0.83 $\pm$ 0.15 & 0.03 $\pm$ 0.02 & 29.84 \\
 &   10 &       0* & 21.39 $\pm$ 0.02 & 0.78 $\pm$ 0.15 & 0.05 $\pm$ 0.03 & 21.51 \\
 &     9 &       3 & 21.81 $\pm$ 0.03 & 0.61 $\pm$ 0.15 & 0.04 $\pm$ 0.03 & 8.16 \\
 &   10 &       1 & 22.57 $\pm$ 0.01 & 0.70 $\pm$ 0.16 & 0.03 $\pm$ 0.02 & 20.60 \\
 &   10/11 &       2/0 & 23.43 $\pm$ 0.09 & 0.98 $\pm$ 0.14 & 0.21 $\pm$ 0.08 & 24.70 \\
\hline
\end{tabular}
\end{table}

\end{document}